\documentclass[final,3p,times]{elsarticle}

\usepackage{yfonts}
\usepackage{graphicx}% Include figure files
\usepackage{dcolumn}% Align table columns on decimal point
\usepackage{bm}% bold math
\usepackage{nicefrac}
\usepackage{amsmath,amsfonts}
\usepackage{romannum}
\usepackage{color}
\usepackage[dvipsnames]{xcolor}
\usepackage{cancel}
\definecolor{darkblue}{RGB}{0,0,196}
\definecolor{darkgreen}{RGB}{0,120,0}
\usepackage[colorlinks=true,linktocpage=true,linkcolor=darkblue,citecolor=red,urlcolor=darkblue]{hyperref}
\usepackage{amssymb}
\usepackage{lipsum}
\usepackage[inkscapearea=page]{svg}
\usepackage{appendix}

\newcommand{\bea}{\begin{eqnarray}}
\newcommand{\eea}{\end{eqnarray}}
\newcommand{\bel}[1]{\begin{eqnarray}\label{#1}}
\newcommand{\eel}{\end{eqnarray}}

\def\LB{\left(}
\def\RB{\right)}
\def\LSB{\left[}
\def\RSB{\right]}
\def\LAB{\langle}
\def\RAB{\rangle}

\newcommand{\EQ}[1]{Eq.~(\ref{#1})}
\newcommand{\EQn}[1]{(\ref{#1})}

\newcommand{\EQSTWO}[2]{Eqs.~(\ref{#1})~and~(\ref{#2})}

\newcommand{\CIT}[1]{Ref.~\citep{#1}} 
\newcommand{\CITn}[1]{\citep{#1}} 

\newcommand{\p}{\partial}

\newcommand{\f}[2]{\frac{#1}{#2}}

\journal{Physics Letters B}

\begin{document}

\begin{frontmatter}

\title{Dynamical constraints on pseudo-gauge transformations}

\author[first]{Zbigniew Drogosz}
\ead{zbigniew.drogosz@alumni.uj.edu.pl}
\affiliation[first]{organization={Institute of Theoretical Physics, Jagiellonian University},
            city={Kraków},
            postcode={30-348}, 
            country={Poland}}
\affiliation[second]{organization={Institute of Nuclear Physics, Polish Academy of Sciences},
            city={Kraków},
            postcode={31-342}, 
            country={Poland}}
\author[first]{Wojciech Florkowski}
\ead{wojciech.florkowski@uj.edu.pl}
\author[first]{Mykhailo Hontarenko}
\ead{mykhailo.hontarenko@student.uj.edu.pl}
\author[second]{Radoslaw Ryblewski}
\ead{radoslaw.ryblewski@ifj.edu.pl}

\begin{abstract}
Classical pseudo-gauge transformations are discussed in the context of hydrodynamic models of heavy-ion collisions. A decomposition of the pseudo-gauge transformation into Lorentz-invariant tensors is made, which allows for better interpretation of its physical consequences. For pseudo-gauge transformations connecting two symmetric energy-momentum tensors, we find that the super-potential $\Phi^{\lambda \mu \nu}$ must obey a conservation law of the form $\partial_\lambda \Phi^{\lambda \mu \nu} = 0$. This equation, referred to below as the STS condition, represents a constraint that is hardly possible to be satisfied for tensors constructed out of the basic hydrodynamic variables such as temperature, baryon chemical potential, and the hydrodynamic flow. However, in a special case of the boost-invariant flow, the STS condition is automatically fulfilled and a non-trivial residual pseudo-gauge transformation defined by a single scalar field is allowed. In this case the bulk and shear viscosity coefficients become pseudo-gauge dependent; however, their specific linear combination appearing in the equations of motion remains pseudo-gauge invariant. This finding provides new insights into the role of pseudo-gauge transformations and pseudo-gauge invariance.
\end{abstract}


\end{frontmatter}

\section{Introduction}
\label{introduction}

The energy-momentum tensor is known to be defined only up to the derivative of an antisymmetric tensor (for a discussion of this point see, for example, \CIT{Coleman:2018mew}). To our knowledge, this non-uniqueness was first addressed by Belinfante in 1939~\citep{belinfante1939spin} and later by Rosenfeld in 1940~\cite{rosenfeld1940energy}. Further analysis of this issue can be found in~\CIT{HEHL197655}, where a simultaneous modification of both the energy-momentum and spin tensors is discussed in depth. The ambiguity in defining these tensors has recently been a topic of active debate, particularly in the context of the proton spin puzzle, where the terms defining these transformations are called boundary terms, as they do not alter the globally conserved quantities~\citep{LEADER2014163}. More recently, the possibility of having and using different versions of the energy-momentum and spin tensors has appeared in the works developing formalism of spin hydrodynamics. In this context, the modifications of the tensors are called the pseudo-gauge transformations (PGT)~\citep{Speranza:2020ilk}. In the following, we use this terminology and refer to pseudo-gauge invariance to describe situations in which certain objects or equations remain unaffected by PGTs.

Pseudo-gauge transformations can be performed at both the classical and quantum levels. At the classical level they lead, in particular, to a change in the energy density, identified with the $T^{00}$ component of the energy-momentum tensor. From the perspective of heavy-ion physics, this result is quite unsettling, as the energy density holds a crucial role in determining the phase transition from ordinary hadronic matter to the quark-gluon plasma (QGP). On the other hand, since the equations of motion (EOM) for conserved currents remain invariant under PGTs, the latter are generally assumed to have no physical significance~\CITn{Coleman:2018mew}. Clearly, such a dichotomy indicates that our concept of energy density is very naive and requires thorough elucidation (that is missing at the moment). 

In quantum field theory (QFT) calculations, some results exhibit a PGT dependence, while others do not~\citep{Becattini:2012pp, Nakayama:2012vs,Fukushima:2020ucl, Li:2020eon, Buzzegoli:2021wlg, Das:2021aar, Weickgenannt:2022jes, Dey:2023hft,Buzzegoli:2024mra}. For example, the quantum calculations of the energy density in a relativistic Bose or Fermi gas are found to be PGT independent, while the fluctuations of the energy in small volumes are PGT dependent, although the latter become PGT independent if the system's volume becomes sufficiently large. The last finding indicates that the issue of the pseudo-gauge invariance may be related to the concept of an appropriate coarse-graining of the system. 

In this work we discuss classical pseudo-gauge transformations, having in mind recent hydrodynamic models of heavy-ion collisions. We perform first a decomposition of the pseudo-gauge transformation into Lorentz-invariant tensors, which allows for a better interpretation of its physical significance. For pseudo-gauge transformations connecting two symmetric energy-momentum tensors, we find that the super-potential $\Phi^{\lambda \mu \nu}$ obeys the conservation law $\p_\lambda \Phi^{\lambda \mu \nu} = 0$ (below dubbed the STS condition, as leading from one {\it symmetric} energy-momentum tensor {\it to} another {\it symmetric} one). In general, this constraint cannot be satisfied by super-potentials constructed out of the basic hydrodynamic variables such as temperature, $T$, baryon chemical potential, $\mu$, and the hydrodynamic flow $u^\mu$. However, in a special case of the boost-invariant flow, the STS condition is automatically fulfilled, and a non-trivial residual pseudo-gauge transformation defined by a single scalar field is allowed.  

This leads us to our main result which shows that although the values of the kinetic coefficients are found to be PGT dependent, the dynamics of the system (its EOM) is PGT invariant. This is so since EOM are determined by the PGT-invariant combination of the shear and bulk viscosity ($\eta$ and $\zeta$). As the independence of the equations of motion is expected on the general grounds, our calculation for the first time  {\it explicitly} shows that this happens with the PGT-modified values of both $\eta$ and $\zeta$.

The paper is organized as follows: In Sec.~\ref{sec:decomp}, we provide general information about the pseudo-gauge transformation and present its explicit decomposition into Lorentz-invariant tensors.  In Sec.~\ref{sec:sts}, we discuss constraints imposed on the PGT in the case where the transformation connects two symmetric energy-momentum tensors. Moreover, we introduce therein the conditions connected with the fact that PGTs are constructed out of the hydrodynamic fields. The Bjorken one-dimensional expansion that allows for a residual PGT is discussed in Sec.~\ref{sec:bjorken}. Explicit expressions for the pseudo-gauge transformations of various components of the energy-momentum tensor are given in~\ref{sec:app}.

{\it{Conventions and notation:}}
Throughout the text we use natural units, $\hbar = c = 1$. The metric tensor is of the ``mostly-minuses'' form, $g_{\mu\nu} = \textrm{diag}(+1,-1,-1,-1)$. For the Levi-Civita tensor $\epsilon^{\mu\nu\alpha\beta}$ we follow the sign convention $\epsilon^{0123} =-\epsilon_{0123} = +1$. The projection operators $\Delta^{\mu \nu }$ and $\Delta^{\mu \nu}_{\alpha \beta} $ are defined by the expressions
%s=
\begin{equation}
\Delta^{\mu \nu } \equiv g^{\mu \nu} -u^\mu u^\nu,   \hspace{0.6cm} \Delta^{\mu \nu}_{\alpha \beta} \equiv \f{1}{2} (\Delta^{\mu}_{\phantom{\mu}{\alpha}} \Delta^{\nu}_{\phantom{\nu}{\beta}}  + \Delta^{\mu}_{\phantom{\nu}{\beta}} \Delta^{\nu}_{\phantom{\nu}{\alpha}}  -\f{2}{3}\Delta^{\mu \nu } \Delta_{\alpha \beta } )\iffalse, \hspace{0,6cm}X^{\LAB \alpha  \RAB } = \Delta^{\alpha \beta}X_\beta \fi.
\end{equation}
The contraction $\Delta^{\mu\nu}_{\alpha\beta}\, A^{\alpha \beta} \equiv  A^{\LAB \alpha \beta \RAB}$ selects the orthogonal (to $u$), symmetric, and traceless part of the tensor $A^{\alpha \beta}$. Round (square) brackets denote symmetrization (antisymmetrization) of Lorentz indices:  $A^{(\mu\nu)} \equiv \frac{1}{2}(A^{\mu\nu}+A^{\nu\mu})$ and $A^{[\mu\nu]} \equiv  \frac{1}{2} (A^{\mu\nu}-A^{\nu\mu})$, except that angular brackets in combination with square ones around a pair of indices denote the
orthogonal asymmetric part
$ A^{\langle [ \alpha \beta ] \rangle} \equiv
\f{1}{2} \LB \Delta^{\mu}{}_{\alpha}\Delta^{\nu}{}_{\beta} - \Delta^{\mu}{}_{\beta}\Delta^{\nu}{}_{\alpha} \RB A^{\alpha \beta}$.
Angular brackets around a single Lorentz index denote orthogonal projection $H^{\alpha \beta \LAB \gamma \RAB \delta ...} \equiv \Delta^{\gamma}_{\phantom{\gamma}{\rho}} H^{\alpha \beta \rho \delta ...}$. Differential operators take precedence over the $\langle \rangle$ symbols. We commonly use the decomposition of a derivative as $\partial_\mu = u_\mu D +\nabla_\mu$ with $D\equiv u_\nu \p^\nu$ and $\nabla_\mu\equiv \Delta_{\mu \nu }\p^\nu$. In addition, we use the expansion scalar $\theta = \nabla \cdot u$.

%%%%%%%%%%%%%%%%%%%%%%%%%%%%%%%%%%%%%%
\section{Lorentz decomposition of the pseudo-gauge transformation}\label{sec:decomp}

The general form of the pseudogauge transformation that does not change the total conserved charges has been discussed in greater detail in~\CIT{HEHL197655} and has the following form for the energy-momentum tensor
\bel{eq:pgt}
T^{\prime \, \mu \nu} = T^{\mu \nu} + \f{1}{2}\p_{\lambda}G^{\lambda \mu \nu} = T^{\mu \nu } + \f{1}{2}\partial_\lambda (\Phi^{\lambda \mu \nu } - \Phi^{\mu \lambda \nu } - \Phi^{\nu \lambda \mu }  ). 
\eel
Here $\Phi^{\lambda \mu \nu }$ denotes an arbitrary rank-3 tensor that is antisymmetric in the last two indices, commonly referred to as the~super-potential. The original and transformed energy-momentum tensors, $T^{\mu \nu}$ and $T^{\prime \, \mu \nu}$, in general contain antisymmetric parts. It is important to observe that both $\Phi^{\lambda \mu \nu } - \Phi^{\mu \lambda \nu }$ and $G^{\lambda \mu \nu}$ are antisymmetric under the exchange $\lambda\leftrightarrow\mu$, hence, if $T^{\mu \nu}$ is conserved, $T^{\prime \,\mu \nu}$ is conserved as well. 

The super-potential can be decomposed into components that are either parallel or transverse to the flow four-vector $u^\mu$. Following the method described in Ref.~\cite{Biswas:2023qsw}, we write
\bel{eq:phi_decompose}
\Phi^{\lambda \mu \nu } =   u^\lambda \mathcal{S}^{\mu \nu}  + (u^\nu \Delta^{\lambda \mu} - u^\mu \Delta^{\lambda \nu})I    + (u^\nu I^{\langle \lambda \mu\rangle }_{(s)} - u^\mu I^{\langle \lambda \nu \rangle }_{(s)}) +  (u^\nu I^{\lambda \mu }_{(a)} - u^\mu I^{\lambda \nu  }_{(a)})   + \Phi^{ \langle \lambda \rangle  \langle \mu \rangle  \langle \nu\rangle},
\eel
where we have used the following definitions:
\bea
\mathcal{S}^{\mu \nu} = u_\alpha \Phi^{\alpha \mu \nu}, \hspace{0.4cm} 
I = \frac{1}{3}I^\mu_{\phantom{\mu}{\mu}}, 
\hspace{0.4cm} I^{\lambda \nu } = -u_\rho \Phi^{\LAB \lambda \RAB \rho \nu}, 
\hspace{0.4cm} \Phi^{\LAB \lambda \RAB  \LAB \mu  \RAB  \LAB \nu \RAB }=
\Delta^{\lambda}_{\phantom{\lambda}{\alpha}} 
\Delta^{\mu}_{\phantom{\mu}{\beta}}
\Delta^{\nu}_{\phantom{\nu}{\gamma}}
\Phi^{\alpha \beta \gamma}.
\eea
Above, we split the tensor $I^{\mu \nu}$ into antisymmetric and symmetric parts ($I^{\mu \nu}_{(a)}$ and $I^{\mu \nu}_{(s)}$, respectively), and from the symmetric part we further extracted the traceless part $I^{\langle \mu \nu \rangle }_{(s)}$. The trace is included in the scalar $I$. Moreover, we decompose the antisymmetric tensor $\mathcal{S}^{\mu \nu}$ as 
\bea
\mathcal{S}^{\mu \nu} =  F^\mu u^\nu - F^\nu u^\mu  + H^{\mu \nu} = F^\mu u^\nu - F^\nu u^\mu + \epsilon^{\mu \nu \alpha \beta }
u_{\alpha } W_{\beta},
\eea
where the four-vectors $F^\mu$ and $W^\mu$ introduced above are orthogonal to the flow vector $u^\mu$ (i.e., $F\cdot u = 0$ and $W \cdot u = 0$).  We note  that the number of independent degrees of freedom on the two sides of~\EQ{eq:phi_decompose} is the same, as discussed in more detail in Ref.~\cite{Biswas:2023qsw} .

In general, the energy-momentum tensor $T^{\mu \nu}$ in Eq.~\EQn{eq:pgt} can be decomposed as follows (likewise for $T^{\prime\mu \nu}$) 
\bel{eq:Tmunudec}
T^{\mu \nu}=\mathcal{E} u^\mu u^\nu -\mathcal{P} \Delta^{\mu \nu}+2 \mathcal{Q}^{(\mu} u^{\nu)}+\mathcal{T}^{\mu \nu}+2 \mathcal{H}^{[\mu} u^{\nu]}+\mathcal{F}^{\mu \nu},
\eel
where the last two terms account for possible antisymmetric contributions. The symmetric parts can be obtained through contractions:  $\mathcal{E} =u_\mu u_\nu T^{\mu \nu},$ $\mathcal{P} = -\f{1}{3}\Delta_{\mu \nu}T^{\mu \nu},$\, $ \mathcal{Q}^\mu= \f{1}{2} \LB \Delta^{\mu}{}_{\alpha}u_{\beta} + \Delta^{\mu}{}_{\beta}u_{\alpha} \RB T^{\alpha \beta}  $, and $\mathcal{T}^{\mu \nu}=T^{\langle\mu \nu\rangle}$, with the standard interpretation of energy-density, isotropic pressure, heat flow, and the shear-stress tensor, respectively (we follow here the notation of~\CIT{Kovtun:2019hdm}). For completeness, we note that the antisymmetric part of \EQ{eq:Tmunudec} is defined in terms of components through the relations $ \mathcal{H}^\mu= \f{1}{2} \LB \Delta^{\mu}{}_{\alpha}u_{\beta} - \Delta^{\mu}{}_{\beta}u_{\alpha} \RB T^{\alpha \beta}$ and $ \mathcal{F}^{\mu\nu}=   T^{\langle[\mu \nu]\rangle}$. 

The pseudo-gauge transformation~\EQ{eq:pgt} induces changes of the coefficients in the decomposition of the energy-momentum tensor~\EQ{eq:Tmunudec}. The explicit form of such changes is given in~\ref{sec:app}.

%%%%%%%%%%%%%%%%%%%%%%%%%%%%%%%%%%%%%
\section{Symmetric energy-momentum tensors and gradient expansion}\label{sec:sts}

In the context of relativistic hydrodynamics, it is interesting to study the case where the initial and final energy-momentum tensors are symmetric. In addition, while considering the hydrodynamic frameworks, one should consider pseudo-gauge transformations constructed out of the hydrodynamic fields such as $T$, $\mu$, and the flow vector $u^\mu$. Furthermore, as long as we restrict our considerations to the case where the transformed energy-momentum tensor does not contain gradients of the order higher than one, the super-potential itself cannot be built from gradients of the hydrodynamic fields. These three conditions very strongly constrain the freedom of pseudo-gauge transformations.

First of all, the condition that both the original and transformed energy-momentum tensors are symmetric leads to the condition~\cite{Speranza:2020ilk}
\bel{eq:STS}
  \p_\lambda  \Phi^{\lambda \mu \nu } = 0,
\eel
which, using~\EQ{eq:phi_decompose}, can be rewritten as
\begin{align}\begin{split}\label{eq:STS1}
\p_\lambda \Phi^{\lambda \mu \nu} &= \theta \mathcal{S}^{\mu \nu} +DH^{\mu \nu}+2  
\LSB F^{[\mu}a^{\nu]} + u^{[\nu}DF^{\mu]} + u^{[\nu} \partial^{\mu]}I + I \p^{[\mu}u^{\nu]}\RSB+ \p_\lambda  \Phi^{ \langle \lambda \rangle  \langle \mu \rangle  \langle \nu\rangle}
  \\ &+ u^{\nu}\p_\lambda \LB I^{\langle \lambda \mu \rangle}_{(s)} +  I^{\lambda \mu }_{(a)} \RB- u^{\mu}\p_\lambda \LB I^{\langle \lambda \nu \rangle}_{(s)} + I^{\lambda \nu }_{(a)} \RB + \LB  I^{\langle \lambda \mu \rangle}_{(s)} +  I^{\lambda \mu }_{(a)}\RB \p_\lambda u^{\nu} -  \LB  I^{\langle \lambda \nu \rangle}_{(s)} +  I^{\lambda \nu }_{(a)}\RB \p_\lambda u^{\mu} =0.
\end{split}\end{align}
We call this equation the STS (symmetric-to-symmetric) condition. An example of the case considered here is the transformation from the GLW to the Belinfante energy-momentum tensor \citep{Florkowski:2018fap}, where the super-potential corresponds to the spin tensor $S^{\lambda, \mu \nu}_{\text{GLW}}$, which is conserved since the respective energy-momentum tensor $T^{\mu \nu}_{\text{GLW}}$ is symmetric (the acronym GLW stands here for the version of the energy-momentum and spin tensors introduced by de Groot, van Leeuwen, and van Weert in \CIT{DeGroot:1980dk}).

The conditions that the pseudo-gauge transformations can be built out of the hydrodynamic variables with the exclusion of gradients further restricts the allowed forms of PGT's. Except one, all terms appearing in the construction of the pseudo-gauge transformation vanish. This is so since we cannot construct the tensors that are orthogonal to $u^\mu$ with the help of just $u^\mu$, with the exception for the projector $\Delta^{\mu \nu}$.~\footnote{General transformation rules for the coefficients appearing in the decomposition \EQn{eq:Tmunudec}  are given in~\ref{sec:app}. In the considered case the vector $F^\mu$ should be proportional to $u^\mu$ and its contribution in \EQn{eq:Glmn} vanishes. Similarly, we cannot construct the tensor $H^{\mu \nu}$ since we do not have any pseudovector at our disposal. Similar although a bit more complicated arguments can be applied to other terms in~\EQn{eq:Glmn}.} As a consequence of this strong assumption, we are left with the form
\bel{eq:Kovtun_transform}
\Phi^{\lambda \mu \nu} = (u^\nu \Delta^{\lambda \mu}  - u^\mu \Delta^{\lambda \nu })I.
\eel
Thus, the STS condition \EQn{eq:STS} means that the following equation should hold
\bel{eq:conservation_Phi}
\p_\lambda \Phi^{\lambda \mu \nu}  = (u^\nu \p^\mu - u^\mu \p^\nu)I + I (\p^\mu u^\nu - \p^\nu u^\mu) = 0.
\eel
Strictly speaking, \EQ{eq:conservation_Phi} represents six equations to be satisfied by one scalar function. Consequently, a pseudo-gauge transformation that maps a symmetric energy-momentum tensor to another symmetric energy-momentum tensor does not exist in relativistic hydrodynamics unless the system (hydrodynamic expansion) possesses additional symmetries.~\footnote{A possible solution to this issue involves introducing additional degrees of freedom into the theory, as in spin hydrodynamics where the four-vector $k^\mu$ and the pseudo-vector $\omega^\mu$ are used \cite{Florkowski:2018fap}. However, herein we consider systems described by $T$, $\mu$, and $u^\mu$ only.}

%%%%%%%%%%%%%%%%%%%%%%%%%%%%%%%%%%%%%%%%%%%%%%%%%%%%%%
\section{Residual PGT for boost-invariant expansion}\label{sec:bjorken}
Equation~\EQn{eq:conservation_Phi} is satisfied for the one-dimensional Bjorken expansion~\cite{Bjorken:1982qr}. In this case, a residual PGT is allowed of the form 
\bel{eq:coefficients}
\mathcal{E}^\prime = \mathcal{E} + I \theta, \hspace{0.6cm} \mathcal{P}^\prime = \mathcal{P} - DI - \f{2}{3}I \theta, \hspace{0.6cm} \mathcal{Q}^{\prime\mu} = \mathcal{Q}^\mu, \hspace{0.6cm}  \mathcal{T}^{\prime\mu \nu} =  \mathcal{T}^{\mu \nu} - I \sigma^{\mu \nu}.
\eel
One can easily check that the presence of the function $I(\tau)$ does not affect the equations of motion.

\subsection{Implementation of the equation of state}

For the sake of simplicity, we shall now consider the case with vanishing baryon chemical potential. Since the PGT changes the energy density, ${\cal E} \to {\cal E}^\prime$, we may interpret it as a change of the effective temperature, $T \to T^\prime$, which is defined by the equations
\bel{eq:TtoTp}
 {\cal E}_{\rm eq}(T^\prime) =
{\cal E}_{\rm eq}(T) + I(T) \theta.
\eel 
Here the function ${\cal E}_{\rm eq}(T)$ is externally given and defines the equation of state of matter under study. Moreover, both $I$ and~$\theta$ should be treated as functions of $T$ (or the proper time $\tau$, with $\theta = 1/\tau$). Hence, the above equations can be used to obtain $T^\prime$ for a given $T(\tau)$. Similarly to \EQn{eq:TtoTp}, we may define the change of pressure
\bel{eq:PtoPp}
 {\cal P}_{\rm eq}(T^\prime) + \Pi^\prime(T^\prime) =
{\cal P}_{\rm eq}(T) + \Pi(T) - D I - \f{2}{3} I \theta.
\eel 
Here the functions $\Pi(T)$ and $\Pi^\prime(T^\prime)$ define the bulk pressure, i.e., an isotropic correction to the equilibrium pressure defined by the functions ${\cal P}_{\rm eq}(T)$ and ${\cal P}_{\rm eq}(T^\prime)$. We stress that the functions $\Pi$ and $\Pi^\prime$ are expected to be different functions of their arguments, and $\Pi^\prime$ does not denote the derivative of $\Pi$. It is convenient to think that with $T^\prime$ determined by \EQ{eq:TtoTp}, one can determine the function $\Pi^\prime$ from \EQ{eq:PtoPp}.

With the parameterizations introduced above, we may write the form of the original energy-momentum tensor as
\bel{eq:Tmunu}
T^{\mu\nu} = {\cal E}(T) u^\mu u^\nu - \left({\cal P}_{\rm eq}(T) + \Pi(T) \right) \Delta^{\mu\nu} + {\cal T}^{\mu\nu},
\eel
and after the PGT as
\bel{eq:Tmunup}
T^{\prime \, \mu\nu} = {\cal E}(T^\prime) u^\mu u^\nu - \left({\cal P}_{\rm eq}(T^\prime) + \Pi^\prime(T^\prime) \right) \Delta^{\mu\nu} + {\cal T}^{\prime \, \mu\nu}.
\eel
In the next step, it is convenient to express the bulk pressures in terms of the bulk viscosity coefficients, and the shear stress tensors in terms of the shear viscosity coefficients,
\bel{eq:zetaeta}
\Pi(T) &=& -\zeta(T) \theta, \quad \Pi^\prime(T^\prime) = -\zeta^\prime(T^\prime) \theta, \\
{\cal T}^{\mu\nu} &=& 2 \eta(T) \sigma^{\mu\nu}, \quad
{\cal T}^{\prime \, \mu\nu} = 2 \eta^\prime(T^\prime) \sigma^{\mu\nu}.
\eel
Here we expect that the functions $\zeta$ and $\zeta^\prime$ are different, similarly as $\eta$ and $\eta^\prime$. These differences manifest dependence of the transport coefficients on PGT.

%%%%%%%%%%%%%%%%%%%%%%%%%%%%%%%%%%%%%%%
\subsection{Hydrodynamic equations}

In our boost-invariant setup with vanishing baryon chemical potential, the hydrodynamic equations are reduced to a single equation of the form
\bel{eq:HE}
\f{d{\cal E}_{\rm eq}(T)}{d\tau} + \frac{{\cal E}_{\rm eq}(T)+{\cal P}_{\rm eq}(T)}{\tau} -  \left[
\f{4}{3} \eta(T) + \zeta(T) \right] \theta^2 = 0,
\eel
and after the PGT to
\bel{eq:HEp}
\f{d{\cal E}_{\rm eq}(T^\prime)}{d\tau} + \frac{{\cal E}_{\rm eq}(T^\prime)+{\cal P}_{\rm eq}(T^\prime)}{\tau} -  \left[
\f{4}{3} \eta^\prime(T^\prime) + \zeta^\prime(T^\prime) \right] \theta^2 = 0.
\eel
Since $T^\prime$ in \EQ{eq:HEp} can be treated as a dummy variable, one can change it to $T$ to obtain the constraint
\bel{eq:C1}
\f{4}{3} \left[\eta^\prime(T)-\eta(T)\right] + \zeta^\prime(T)-\zeta(T) = 0.
\eel
Substituting \EQ{eq:HE} in \EQ{eq:HEp} leads to another condition
\bel{eq:C2}
\eta^\prime(T^\prime) = \eta(T) - \f{I(T)}{2} .
\eel
The last two equations imply that both shear and bulk viscosities may change under PGT, however, their combination defined by \EQ{eq:C1} should remain unchanged. We note that \EQ{eq:C2} directly follows from \EQ{eq:coefficients}.

%%%%%%%%%%%%%%%%%%%%%%%%%%%%%%%%%%%%%%%%%%%%%%%
\subsection{Conformal case}

For conformal systems, \EQ{eq:C1} is quite restrictive. In this case, the trace of the energy-momentum tensor vanishes implying that both $\zeta$ and $\zeta^\prime$ vanish. Moreover, all thermodynamic functions scale with temperature, hence we have: ${\cal E}_{\rm eq}(T) = a T^4$, ${\cal P}_{\rm eq}(T) = (a/3) T^4$, and the entropy density ${\cal S}_{\rm eq}(T) = (4 a/3) T^3$, where $a$ is a constant. The shear viscosity can be written as a constant ${\bar \eta}$ multiplying the entropy density, $\eta(T) = (4 {\bar \eta} a/3) T^3$ (for strongly interacting matter ${\bar \eta} = 1/(4\pi)$). From dimensional analysis we also have $I(T) = b T^3$, where $b$ is yet another constant. In this case \EQ{eq:C2} gives
\bel{eq:C21}
T^\prime = T \left(1-\f{3b}{8a{\bar \eta}} \right)^{1/3}.
\eel
On the other hand, directly form \EQ{eq:TtoTp} we obtain $T^{\prime \, 4} = T^4 + b T^3/(2\tau)$, which contradicts the scaling~\EQn{eq:C21}---from the hydrodynamic equation \EQn{eq:HE} we may conclude that $T$ does not decrease as $1/\tau$. 

%%%%%%%%%%%%%%%%%%%%%%%%%%%%%%%%%%%%%%%%%%%%%%%
\subsection{Non-conformal case}

In the non-conformal case, we do not have any constraints restricting the trace of the energy-momentum tensor and the bulk viscosity coefficient. As a direct consequence of Eqs.~\EQn{eq:TtoTp}, \EQn{eq:PtoPp}, \EQn{eq:C1}, and \EQn{eq:C2}, we obtain a differential equation for the function $I(\tau)$, namely
\bel{eq:I}
\f{dI}{d\tau}+\f{I}{3\tau} + A_{\rm eq}(T^\prime) - A_{\rm eq}(T) = 
\f{4}{3\tau} \left[ \eta(T^\prime)-\eta(T) \right] +
\f{1}{\tau} \left[ \zeta(T^\prime)-\zeta(T)\right],
\eel
where $A_{\rm eq}(T)$ describes the trace anomaly, $A_{\rm eq}(T) = {\cal P}_{\rm eq}(T)- (1/3) {\cal E}_{\rm eq}(T)$. Equations~\EQn{eq:TtoTp} and \EQn{eq:I} can be used to determine the time evolution of the functions $T^\prime(\tau)$ and $I(\tau)$. Then, \EQ{eq:C2} serves to obtain the function $\eta^\prime$, and, finally, \EQ{eq:C1} can be used to obtain the function $\zeta^\prime$. 

\begin{figure}[t]
\hspace{1.5cm}
\includegraphics[width=0.7 \textwidth]{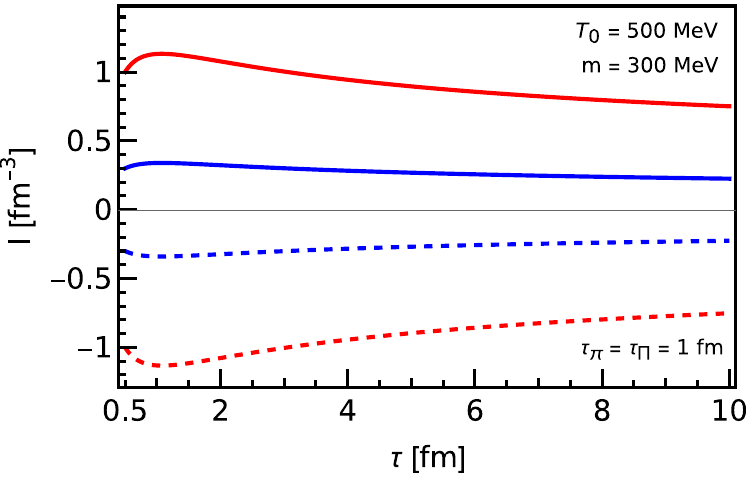}
\caption{Proper time dependence of the function $I(\tau)$ for different initial conditions. Calculations were performed for the hadron gas model described in \CIT{Denicol:2014mca}.}\label{plot}
\end{figure}

We constructed numerical solutions of \EQSTWO{eq:TtoTp}{eq:I} using the equation of state describing relativistic gas of classical massive particles, and the forms of kinetic coefficients were taken from~\CIT{Denicol:2014mca}. The calculations were done with the initial time $\tau_0 = 0.5$~fm, the initial temperature $T_0 = T(\tau_0) = 500$~MeV, the particle mass $m = 300$~MeV, and the shear and bulk relaxation times $\tau_\pi = \tau_\Pi = 1$~fm. Our results are presented in Fig.~\ref{plot} for four different initial values of the function $I$: $I(\tau_0) = \pm 1$~fm$^{-3}$ and $I(\tau_0) = \pm 0.3$~fm$^{-3}$. Our results show that the function $I$ decreases slowly with time, which is expected since the function $I=0$ is a solution of \EQ{eq:I}.

%%%%%%%%%%%%%%%%%%%%%%%%%%%%%%%%%%%%%%%
\section{Summary and conclusions}
In this work, we have examined the pseudo-gauge transformations using a Lorentz-invariant decomposition of the super-potential. By assuming that the energy-momentum tensors are symmetric (before and after the PGT), we have derived the STS condition, which must be satisfied by the super-potential. This result shows that PGT cannot be applied to every system owing to the limited number of degrees of freedom. However, a particular class of transformations, called the residual PGTs, are allowed for systems exhibiting very special symmetry constraints. An example of such a system is the one-dimensional Bjorken model. In this case
the bulk and shear viscosity coefficients become PGT dependent, however, their specific linear combination
appearing in the equations of motion remains pseudo-gauge invariant. This result indicates an interesting phenomenon: a PGT dependence of one of the kinetic coefficients might be irrelevant for the whole dynamics of the system, as it is compensated by the PGT changes of other coefficients.

\section*{Acknowledgments}
It is a pleasure to thank Valeriya Mykhaylova for useful discussions. This work was supported in part by the Polish National Science Centre (NCN) Grants No. 2022/47/B/ST2/01372 and No.
2018/30/E/ST2/00432.

\appendix

\section{Pseudo-gauge transformation rules for the energy-momentum coefficients} \label{sec:app}

With the parametrization~\EQn{eq:phi_decompose}, we obtain
\begin{align}\begin{split}
G^{\lambda \mu \nu} &=  u^\lambda H^{\mu \nu} - u^\mu H^{\lambda \nu} - u^\nu H^{\lambda \mu}
   +2 \LSB  I (u^\lambda \Delta^{\mu \nu } -   u^\mu \Delta^{\nu \lambda})  + u^\lambda I^{\LAB \mu\nu  \RAB}_{(s)}- u^\mu I^{\LAB \lambda \nu \RAB}_{(s)}  +  u^\nu I^{\lambda \mu}_{(a)} \RSB \\ &+ 2 \LB    u^\nu u^\lambda F^\mu - u^\mu u^\nu F^\lambda \RB + 
   \Phi^{ \langle \lambda \rangle  \langle \mu \rangle  \langle \nu\rangle}+ \Phi^{ \langle \mu \rangle  \langle \nu\rangle \langle \lambda \rangle }+ \Phi^{ \langle \nu \rangle \langle \mu \rangle \langle \lambda \rangle },
\label{eq:Glmn}
\end{split}\end{align}
which leads to the following transformation rule for the energy-momentum tensor
\begin{align}\begin{split}
T^{\prime\mu \nu} &= T^{\mu \nu}+ (I \theta  - {\theta}_F) u^\mu u^\nu  +  (I \theta+    D I)\Delta^{\mu \nu}  - I u^\nu a^\mu  -  u^\mu \nabla^\nu I  -  I \nabla^\nu u^\mu +  \theta F^\mu u^\nu  + u^\nu D F^\mu \\ 
&+ F^\mu a^\nu -   2u^{(\mu} {D}_F u^{\nu)} + \f{1}{2}\theta H^{\mu \nu} + \f{1}{2}DH^{\mu \nu}- u^{(\mu} h^{\nu)} - D_H^{(\mu} u^{\nu)}+  \theta I^{\langle \mu \nu \rangle}_{(s)} + DI^{\langle \mu \nu \rangle}_{(s)} \\ &+ u^\nu \p_\lambda I^{ \lambda \mu }_{(a)} + I^{ \lambda \mu }_{(a)} \p_\lambda u^\nu - u^\mu \p_\lambda  I^{ \langle \lambda \nu \rangle}_{(s)} -  I^{ \langle \lambda \nu \rangle}_{(s)} \p_\lambda u^\mu
+ \f{1}{2}\p_\lambda \LB \Phi^{ \langle \lambda \rangle  \langle \mu \rangle  \langle \nu\rangle} + \Phi^{  \langle \mu \rangle  \langle \nu\rangle \langle \lambda \rangle } + \Phi^{  \langle \nu \rangle  \langle \mu \rangle \langle \lambda \rangle  }\RB.
\label{eq:pgt2}
\end{split}\end{align}
Here we have introduced the following notation and definitions
\begin{gather}
\nabla \cdot  F = {\theta}_F,  \hspace{0.4cm} \p_{\mu}H^{\mu \nu} = h^\nu,\\
 Du^\nu = a^\nu, \hspace{0.4cm} F^\mu \p_{\mu} = {D}_F, \hspace{0.4cm} H^{\lambda \nu}\p_{\lambda} = D_H^\nu.
\end{gather}
Based on Eq.~\EQn{eq:pgt2}, the symmetric terms in the decomposition~\EQn{eq:Tmunudec} transform according to the rules:
\begin{equation}
\mathcal{E'} = \mathcal{E} + I \theta  - \nabla \cdot F-\theta_w +  I^{\LAB  \alpha \beta \RAB }_{(s)} \sigma_{\alpha \beta}- I^{ \alpha\beta}_{(a)}   \Omega_{\alpha\beta} \,,
\end{equation}
\begin{equation}
\mathcal{P}'=\mathcal{P} - \f{1}{3} \LB 2 I \theta + 3D I + F \cdot a + \theta_w -  I^{\LAB  \alpha \beta \RAB }_{(s)} \sigma_{\alpha \beta} + I^{ \alpha\beta}_{(a)}   \Omega_{\alpha\beta}  + \p_\alpha \Phi^{ \langle \beta \rangle \phantom{\langle \beta \rangle} {\langle \alpha
 \rangle} }_{\phantom{ \langle \beta \rangle}{\langle \beta \rangle}} \RB \,,
\end{equation}
\begin{align}\begin{split}
\mathcal{Q}'^\mu &= \mathcal{Q}^\mu + \f{1}{2} \LSB - I a^\mu + DF^\mu + \theta F^\mu - 2 D_F u^\mu - \nabla^\mu I  + (F \cdot a) \, u^\mu  \RSB + u_\nu D I^{\langle \mu \nu \rangle}_{(s)} - \f{1}{2}u^\mu \LB I^{\LAB  \alpha \beta \RAB }_{(s)} \sigma_{\alpha \beta} -  I^{ \alpha\beta}_{(a)}   \Omega_{\alpha\beta}\RB   \\ &- \f{1}{2} \p_\lambda \LB I^{\langle \mu \lambda \rangle}_{(s)} + I^{\mu \lambda}_{(a)} \RB 
-
\f{1}{2} \LB 
\Phi^{\langle \mu \rangle\langle \nu \rangle\langle \lambda \rangle}
+ 
\Phi^{\langle \nu \rangle\langle \mu \rangle\langle \lambda \rangle} \RB \p_\lambda u_\nu  
+ \frac{1}{2} \epsilon^{\langle \mu \rangle}_{\phantom{\langle \mu \rangle}{\alpha\beta\gamma}} (W^\alpha \p^\beta u^\gamma-u^\alpha \p^\beta W^\gamma) \, ,
\end{split}\end{align}
\begin{align}\begin{split}  
\mathcal{T}^{'\mu \nu}&= \mathcal{T}^{\mu \nu} - I \sigma^{\mu \nu}+ F^{\langle \mu} a^{\nu\rangle}+D I^{\LAB \mu\nu \RAB }_{(s)}+I^{\LAB \mu\nu \RAB }_{(s)} \theta- I^{\LAB \langle \mu \lambda \RAB }_{(s)} \nabla_\lambda u^{\nu\rangle} + 2 u^{(\mu} I^{\LAB \nu )\lambda \RAB }_{(s)} a_{\lambda}+  I^{\lambda \langle\mu }_{(a)} \nabla_\lambda u^{\nu\rangle} \\ 
&+ \epsilon^{\alpha \beta \lambda \langle \mu} W_\alpha u_\beta \p_\lambda u^{\nu \rangle} + \p_\lambda \Phi^{\langle  \mu  \nu  \rangle \langle \lambda \rangle} \, ,
\end{split}\end{align}
where we defined one more scalar $ \theta_w = u \cdot h =  \epsilon^{\mu \nu \alpha \beta } W_\mu u_\nu \nabla_\alpha u_\beta$, the transverse traceless symmetric tensor (shear flow tensor) $\sigma_{\alpha\beta}=\p_{\LAB\alpha} u_{\beta\RAB}$, and the antisymmetric orthogonal tensor $\Omega_{\alpha\beta}\equiv\p_{\langle[\alpha} u_{\beta]\rangle}$.

\bibliographystyle{unsrt}

\end{document}